\documentclass{article}
\usepackage{spconf,amsmath,amsfonts,amssymb,graphicx}
\usepackage[linesnumbered,ruled,vlined]{algorithm2e}
\usepackage{array}
\usepackage{multirow}
\usepackage[romanian]{babel}

\title{TRANSFORMER-BASED ONLINE SPEECH RECOGNITION WITH DECODER-END ADAPTIVE COMPUTATION STEPS}

\name{Mohan Li, C\u{a}t\u{a}lin Zoril\u{a} and Rama Doddipatla}
\address{Toshiba Cambridge Research Laboratory, Cambridge, UK}

\begin{document}
\ninept
\maketitle

\begin{abstract}
Transformer-based end-to-end (E2E) automatic speech recognition (ASR) systems have recently gained wide popularity, and are shown to outperform E2E models based on recurrent structures on a number of ASR tasks. However, like other E2E models, Transformer ASR also requires the full input sequence for calculating the attentions on both encoder and decoder, leading to increased latency and posing a challenge for online ASR. The paper proposes Decoder-end Adaptive Computation Steps (DACS) algorithm to address the issue of latency and facilitate online ASR. The proposed algorithm streams the decoding of Transformer ASR by triggering an output after the confidence acquired from the encoder states reaches a certain threshold. Unlike other monotonic attention mechanisms that risk visiting the entire encoder states for each output step, the paper introduces a maximum look-ahead step into the DACS algorithm to prevent from reaching the end of speech too fast. A Chunkwise encoder is adopted in our system to handle real-time speech inputs. The proposed online Transformer ASR system has been evaluated on Wall Street Journal (WSJ) and AIShell-1 datasets, yielding 5.5\% word error rate (WER) and 7.1\% character error rate (CER) respectively, with only a minor decay in performance when compared to the offline systems.
\end{abstract}


\section{Introduction}
\label{sec:intro}

End-to-end (E2E) architectures have noticeably simplified the efforts of modelling and decoding in automatic speech recognition (ASR) systems. The acoustic model (AM), pronunciation lexicon and language model (LM) that constitute a conventional hybrid system are implicitly integrated into a single neural network, which directly transcribes speech into text via a forward propagation of the model. So far, various E2E modelling techniques have been studied, including Connectionist Temporal Classification (CTC) \cite{graves2006,graves2014}, Recurrent Neural Network Transducer (RNN-T) \cite{graves2012}, and attention-based encoder-decoder frameworks \cite{chorowski2015,chan2016}. However, as most of the E2E model trainings are purely data-driven, the performance is highly dependent on the scale of training data, thus vanilla E2E models were previously believed to achieve comparable results with the hybrid systems only when trained on large datasets. Nevertheless, the gap in performance gradually reduced with more advanced modelling strategies, represented by hybrid CTC/attention architecture \cite{kim2017,watanabe2017}, where the advantages of both structures are utilised in the multi-objective learning. Moreover, externally trained RNN LMs are augmented to E2E models with either rescoring \cite{bahdanau2016,hori2017} or fusion techniques \cite{sriram2018,shan2019}, which effectively push the limits of knowledge acquired in E2E training. Generally, the simplicity in construction and competitive accuracy have made E2E models a strong alternative for building state-of-the-art ASR systems.

Recently, Transformer \cite{vaswani2017} models have been explored for ASR \cite{dong2018}, after the overwhelming success achieved in Neural Language Processing (NLP), and this is no doubt another boost for E2E ASR systems, as better results are reported on a series of open-sourced datasets \cite{dong2019,zhao2019,karita2019}. In contrast with RNN-based architectures, Transformer processes the input sequence completely relying on the self-attention mechanism, by which the dependencies between each pair of elements in the sequence can be equally captured regardless of the distance. Such a property is believed critical to the language modelling in ASR systems. Transformer employs the encoder-decoder framework similar to previous attention based architectures, where the encoder and decoder both consist of stacked self-attention networks (SAN), and are bridged using the cross-attention mechanism. 

Transformers, though have shown to perform better than recurrent attention-based models, still encounter similar problems as bidirectional RNN structures with respect to online speech recognition, where: a) the whole utterance is requested prior to encoding, as the self-attention is calculated for all the positions in the input sequence; b) normalisation is performed over all the encoder states at each output step, which is not only computationally expensive but is a growing concern for memory consumption with increase in input length. The above issues have been extensively studied in RNN-based E2E models. A first attempt to tackle these issues is to slide a window with fixed length and stride on the utterance, restricting the number of frames fed to the encoder at a time \cite{chorowski2015}. Reducing the context might have minimal impact on encoding as the acoustic features tend to rely on local dependencies, but the decoding could be severely destabilised when attentions fall outside the window, which remains uncontrollable due to varying speaking rates or long pauses in speech. A later solution avoids this drawback by predicting the central position of the next alignment \cite{tjandra2017}, where windowing is no longer needed as the predictions are also sequentially made along the encoder states and the decoder is not involved. Similar ideas are presented in \cite{luong2015,hou2017}. So far, the most popular online decoding strategies are based on hard attention mechanism, epitomised by Hard Monotonic Attention (HMA) \cite{luong2017}, where the computation of attentions is formulated as a stochastic (Bernoulli) process, and the output is triggered once the attending probability at a certain timestep is sampled as 1. Monotonic Chunkwise Attention (MoChA) \cite{chiu2018} is later proposed to adapt to the flexibility of speech-to-text-alignment by applying a second-pass soft attention on top of a small chunk, preceding the endpoint detected from the first-pass hard attention. Other variants like Stable Monotonic Chunkwise Attention (sMoChA) \cite{miao2019} and Monotonic Truncated Attention (MTA) \cite{miao2020a} are further developed towards easing the difficulties in model training for online ASR.

Hard attention techniques have also been explored for Transformer ASR in \cite{tsunoo2019,miao2020b,moritz2020}. However, the limits of Bernoulli-based methods are amplified owing to the use of multi-head attention mechanism. Firstly, it is often noticed that the heads focus on different parts of the input, resulting in ambiguous attended timesteps even for a single decoder layer. Secondly, for a good number of heads, the output can never be triggered until the end of speech is reached, which makes the decoding far less online. The above issues encountered in online Transformer ASR are addressed in this paper by proposing a Decoder-end Adaptive Computation Steps (DACS) algorithm that transforms the stochastic process into an accumulation of attention confidence and provides an effective control towards the latency in decoding.

The rest of this paper is structured as follows: Section 2 presents the architecture of Transformer ASR and the streaming methods proposed in literature. Section 3 elaborates the DACS algorithm and integrates it to the Transformer ASR system. Experimental results are presented in Section 4. Finally, conclusions are drawn in Section 5.

\section{Transformer for online ASR}
\label{sec:related}

The baseline architecture of Transformer ASR is derived from \cite{vaswani2017}, which consists of an encoder and a decoder, with both comprising stacked SAN layers. Convolutional neural network (CNN) layers are commonly prepended to the encoder, aiming to extract better acoustic features and perform subsampling on the input frames. Each SAN of the encoder has two sub-layers, known as the multi-head self-attention mechanism and the position-wise feed-forward network. Residual connections are placed around both the modules, and layer normalisation \cite{ba2016} is applied before each sub-layer rather than after. The SANs of the decoder also include a third sub-layer for performing multi-head cross-attention between the states of encoder and decoder. As in encoder, the sub-layers in decoder also have residual connections and layer normalisation. To model the sequence order information that is absent in the self-attention mechanism, positional encodings are added to the CNN output on the encoder side, as well as the token embeddings on the decoder side.

Specifically, Transformer employs Scaled Dot-Product Attention to map a query with a series of key-value pairs to an output by:
\begin{equation} \mathrm{Attention}(\mathbf{Q},\mathbf{K},\mathbf{V}) = \mathrm{softmax}(\frac{\mathbf{Q} \mathbf{K}^T} {\sqrt{d_k}}) \mathbf{V}, \end{equation}
where $\mathbf{Q} \in \mathbb{R} ^{L\times d_k}$ and $\mathbf{K,V} \in \mathbb{R} ^{T\times d_k}$ denote queries, keys and values in the matrix form, $L$ and $T$ are the number of queries and key-value pairs, and $d_k$ is the dimension of representation. In the context of ASR, for self-attention sub-layers, $\mathbf{Q}$, $\mathbf{K}$ and $\mathbf{V}$ are identical as the states of either encoder or decoder. As for cross-attention sub-layers, $\mathbf{Q}$ is the decoder states and $\mathbf{K}$, $\mathbf{V}$ are identical as the encoder states, where the term inside the softmax function in eq. (1) is referred to as attention energy $\mathbf{E} \in \mathbb{R} ^{L\times T}$.

At each SAN layer, multiple heads are utilised to diverge the attentions into distinct sub-spaces, allowing the model to jointly obtain information from different positions:
\begin{equation} \mathrm{MultiHead}(\mathbf{Q},\mathbf{K},\mathbf{V}) = \mathrm{Concat}(\mathrm{head}_1,...,\mathrm{head}_H) \mathbf{W}^O, \end{equation}
\begin{equation} \mathrm{where\;head}_h = \mathrm{Attention}(\mathbf{QW}^Q_h,\mathbf{KW}^K_h,\mathbf{VW}^V_h), \end{equation}
where $\mathbf{W}^{Q,K,V}_h \in \mathbb{R} ^{d_k \times d_k}$ and $\mathbf{W}^O \in \mathbb{R} ^{d_m \times d_m}$ are the parameters of projection matrices, $H$ is the number of attention heads, and $d_m = H\times d_k$.

For Transformer ASR to operate in online mode, some modifications should be made to the cross-attention sub-layers in the decoder. As shown in eq. (1), the softmax is performed on the entire rows of $\mathbf{K}$, assuming that attentions are distributed across the full encoder states. Hard attention mechanisms circumvent this problem by encouraging attentions to concentrate on single encoder states, so as to spare the need of global normalisation. To make the discussion easier for the reader, a step wise notation is employed instead of matrix notation for all the equations presented beyond this point.

In HMA \cite{luong2017}, at the $i^{th}$ output step, a monotonic attention (MA) energy $e_{i,j}$ is calculated for each encoding timestep $j$ based on the last decoder state $q_{i-1}$ and the encoder state $k_j$. Then $e_{i,j}$ is passed to a sigmoid unit to produce an attending probability $p_{i,j}$, from which a decision $z_{i,j}$ is sampled to indicate whether the timestep $j$ is attended or not. The process is formulated as:
\begin{equation} e_{i,j} = \frac {q_{i-1} k_j^T} {\sqrt{d_k}}, \end{equation}
\begin{equation} p_{i,j} = \mathrm{sigmoid}(e_{i,j}), \end{equation}
\begin{equation} z_{i,j} = \mathrm{Bernoulli}(p_{i,j}). \end{equation}
Sigmoid is regarded an effective replacement to the softmax that rescales the energy within the range (0, 1). Due to the involvement of sampling, an expectation form of $p_{i,j}$ is actually used in the HMA model training in order to enable backpropagation, which is defined as:
\begin{equation} \alpha_{i,j} = p_{i,j} \sum_{m=1}^{j} ( \alpha_{i-1,m} \prod_{l=m}^{j-1} ( 1 - p_{i,l} ) ), \end{equation}
where series \{$\alpha_{i,.}$\} will serve as the attention weights in working out the context vector $c_i$:
\begin{equation} c_i = \sum_{j=1}^T \alpha_{i,j} v_j, \end{equation}
Note that for HMA inference, the output is not triggered until a $p_{i,j}>0.5$ is met at timestep $t_i$, and $c_i$ is directly assigned as $v_{t_i}$. Though HMA is sufficient to stream the Transformer ASR, the main drawback is that the use of a single encoder state in calculation of the context vector can be unreliable.

MoChA \cite{chiu2018} enhances HMA by performing another soft attention on top of the encoder states, which facilitates the handling of non-monotonic alignments. The chunkwise energy $u_{i,n}$ is estimated within a $w$-size window ending at $t_i$:
\begin{equation} u_{i,n} = \frac {q_{i-1} k_n^T} {\sqrt{d_k}}, \end{equation}
where $n$ denotes the timestep index in the window. Again, the expectation form of $u_{i,n}$ that also involves $a_{i,j}$ is adopted during training:
\begin{equation} \beta_{i,j} = \sum_{n=j}^{j+w-1} \frac{\alpha_{i,j} \exp(u_{i,n})} {\sum_{l=n-w+1}^{n} \exp(u_{i,l})}. \end{equation}
Now in MoChA, the calculation of $c_i$ is based on \{$\beta_{i,.}$\}:
\begin{equation} c_i = \sum_{j=1}^T \beta_{i,j} v_j, \end{equation}
where \{$\beta_{i,.}$\} are substituted by chunk-normalised \{$u_{i,.}$\} at the inference stage. MoChA has shown good advantage over HMA in terms of attention flexibility, but the difficulty of training remains a big issue as reflected in eq. (7). It is obvious that once \{$\alpha_{i,.}$\} tend to zero for a certain output step, the weak attentions would inevitably pass down to all the following steps \cite{tsunoo2019}, which results in vanished gradients for backpropagation. 

To remedy this, sMoChA \cite{miao2019} is proposed to simplify eq. (7) to:
\begin{equation} \alpha_{i,j} = p_{i,j} \prod_{l=1}^{j-1} ( 1 - p_{i,l} ). \end{equation}
The cost of training is notably cut down by the drop of the recursive term $\alpha_{i-1,j}$ , and the likelihood of attentions is maximised independently for each output step $i$. Moreover, sMoChA may have access to more historical encoder states rather than a small chunk, which is claimed to alleviate the mismatch in attention's receptive scope between training and inference. 

MTA \cite{miao2019} further simplifies the workflow of attention computation, in which the production of attention weights and context vector has been unified for both training and inference, using eq. (12) and (8) respectively. This not only speeds up the model training, but also expands the scope of attention to all the encoding timesteps before the current truncating point. In addition, MTA has so far given the best ASR performance among various hard attention mechanisms according to \cite{miao2020c}.


\section{Online Transformer ASR with decoder-end Adaptive Computation steps}
\label{sec:DACS}

When applying the aforementioned online attention mechanisms in Transformer ASR, the major challenge is to control the latency during inference. Due to the different behaviours of MA heads in Transformer decoder, some of them may produce attending probabilities \{$p_{i,.}$\} that are all below the threshold throughout an output step, which means the end of speech could be easily reached at the early stage of decoding. A possible solution is to convert the Bernoulli process based on $p_{i,j}$ into a more stable accumulation of the attention weights, and the process is halted once the accumulation exceeds a certain threshold. A forced truncation is also applied when the threshold is not reached beyond a fixed number of timesteps, in order to contain the delay in decoding. This leads to the proposed Decoder-end Adaptive Computation Steps (DACS) algorithm in this work.

Adaptive Computation Steps (ACS) algorithm is proposed in \cite{li2019}, and has been applied to the encoder of both recurrent and SAN-based architectures \cite{dong2020}. To dynamically decide the number of frames that are needed to produce an output, a halting layer is imposed on top of the encoder, which consists of an 1-D CNN, a linear vector-to-scalar projector and a sigmoid unit. Halting probabilities are sequentially generated along the encoder states, and the computation is stopped once the accumulation of these probabilities (confidence) reaches a certain threshold. The main weakness of ACS is that the performance is highly sensitive to the output unit. The units with vague acoustic boundaries like English letters or punctuation marks pose to be challenging for ACS \cite{li2019}. 

In this paper, we alter the implementation of ACS into Transformer's self-attention decoder (SAD) by applying it to the cross-attention sub-layers. Instead of relying exclusively on the encoding information, the halting confidence is now jointly estimated from the encoder-decoder correlation, which is more in line with the concept of attention while having the advantage of being computed online.

The workflow of DACS for the MA $head_h$ in the $l^{th}$ SAD layer is elaborated as follows. At output step $i$, similar to eq. (4), an MA energy $e_{i,j}$ is computed for each encoding timestep $j$, given the encoder state $k_j$ and decoder state $q_{i-1}$:
\begin{equation} e_{i,j} = \frac {q_{i-1} k_{j}^T} {\sqrt{d_k}}, \end{equation}
which is then passed to a sigmoid unit to generate the halting probability $p_{i,j}$:
\begin{equation} p_{i,j} = \mathrm{Sigmoid}(e_{i,j}). \end{equation}
Note that eq. (13) differs from the monotonic energy function used in MoChA \cite{tsunoo2019} by removing the scaling and bias terms, together with the norm of $q_{i}$. More particularly, Gaussian noise is not introduced to $e_{i,j}$ during training as the discreteness of $p_{i,j}$ is not encouraged anymore in DACS. 

From $j=1$, we keep inspecting the accumulation of $p_{i,j}$, and halt the computation as soon as the cumulative sum is greater than 1. However, as mentioned above, $head_h$ may provide constant weak energies for all $j$, resulting in very small values for \{$p_{i,.}$\}, which deters halting for the entire utterance. This is overcome by imposing a maximum look-ahead step $M$ to truncate the inspection at $t_{i-1}+M$, where $t_{i-1}$ is the halting position recorded at the last output step $i-1$.The above process can be encapsulated by defining an adaptive computation steps $N_{i}$ as:
\begin{equation} N_{i} = \mathrm{min} \Bigg\{ \mathrm{min} \Bigg\{ n: \sum_{j=1}^n p_{i,j} > 1 \Bigg\}, M \Bigg\}, \end{equation}
which represents the number of encoder states that $head_h$ requires for generating the current output.

As the output of $head_h$, a context vector is calculated using the halting probabilities $\{ p_{i,.} \}$ as weights, similar to eq. (8) and (11):
\begin{equation} c_{i} = \sum_{j=1}^{N_i} p_{i,j} v_{j}, \end{equation}
where $v_{j}$ is identical to the encoder state $k_{j}$. It is observed that the handling of $p_{i,j}$'s remainder \cite{graves2016,li2019} has been ruled out in DACS, while in the original ACS, halting probability of the last computation step $t_i$ (= $t_{i-1}+N_{i}$) should be trimmed so as to make $\{ p_{i,.} \}$ sum up to 1. Yet two major benefits are brought to DACS by the use of untrimmed $\{p_{i,.}\}$ as: (1) it cuts down the potential loss of attention in such a case that the value of $p_{i,t_i}$ is not small, whereas the preceding accumulation has already approached 1, and (2) it accelerates the model training as there's no need to revisit the values of $p_{i,t_i}$ in the matrix.

Furthermore, to coordinate the decoding stages of DACS-based SAD layers, the halting positions given by different MA heads must be carefully synchronised at every output step. Here, we use the unified halting position for the whole SAD despite that unequal $N_i$ might be consumed by separate layers. Imbalanced receptive scopes inside the model are removed via forcing all heads to proceed with the same stride along the encoder states. To achieve this, halting position of output step $i$ is set to the furthest timestep reached by any heads in the structure. A complete illustration of the DACS inference is shown in Algorithm 1.

\begin{algorithm}[t]
\DontPrintSemicolon
\SetAlgoLined
\KwIn{encoder states $\mathbf{k}(\mathbf{v})$, decoder states $\mathbf{q}$, length $T$, maximum look-ahead step $M$, number of heads $H$, number of decoder layers $N_d$. }
 \textbf{Initialization:} $y_0=\langle sos \rangle$, $t_0=0$ \;
 \While{$y_{i-1} \neq \langle eos \rangle$} {
   $t_i=t_{i-1}$ \;
   \For{$l = 1$ \textbf{to} $N_d$} {
      $t^l_i=t_{i}$ \;
      \For{$h = 1$ \textbf{to} $H$} {
        $acc^{h,l}_i=0$ \;
        \For{$j = 1$ \textbf{to} \rm{min}($t_{i-1}+M,T$)} {
            $p^{h,l}_{i,j}=\mathrm{sigmoid}(\frac{q^{h,l}_i k^{T}_j}{\sqrt{d_k}})$ \;
            $acc^{h,l}_i \mathrel{{+}{=}} p^{h,l}_{i,j}$ \;
            \If{$acc^{h,l}_i > 1$} {
                break \;
                }
            }
            $c^{h,l}_i=\sum^j_{m=1}p^{h,l}_{i,m}v^{h,l}_{m}$ \;
            
            $t^l_i=\mathrm{max}(t^l_i,j)$ \;
        }
        $c^l_i=\mathrm{Concat}(c^{1,l}_i,...,c^{H,l}_i)$ \;
        $t_i=\mathrm{max}(t_i,t^l_i)$ \;
    }
    $i \mathrel{{+}{=}} 1$ \;
  }
 \caption{DACS Inference for Transformer ASR}
\end{algorithm}

Finally, we rewrite eq. (13) to (16) in the matrix format and generalise them for all MA heads:
\begin{equation} \mathbf{P} = \mathrm{sigmoid} ( \frac {\mathbf{Q} \mathbf{K}^T} {\sqrt{d_k}} ), \end{equation}
\begin{equation} \mathbf{M} = \mathrm{ShiftRight} ( \mathrm{cumsum} ( \mathbf{P} ) > 1 ) ), \end{equation}
\begin{equation} \mathrm{DACS}(\mathbf{Q},\mathbf{K},\mathbf{V}) = \mathbf{M} \odot \mathbf{P} \mathbf{V}, \end{equation}
where $\mathbf{P}, \mathbf{M} \in \mathbb{R} ^{L \times T}$ are the halting probability matrix and the corresponding halting mask. In eq. (18), the cumulative summation function $\mathrm{cumsum}(\mathbf{x}) = [x_1,x_1+x_2,x_1+x_2+x_3,...,\sum_{k=1}^{|\mathbf{x}|}x_k]$, and the right-shift operator $\mathrm{ShiftRight}(\mathbf{x}) = [0,x_1,x_2,...,x_{|\mathbf{x}|-1}]$ are both applied to the rows of $\mathbf{P}$. The layer-wise output of DACS is a combination of all heads like what is done in eq. (2).

\section{Experiments}
\label{ssec:exp}

\subsection{Experimental setup}
\label{ssec:expsetup}

The proposed DACS-based Transformer ASR is evaluated on two datasets: English Wall Street Journal (WSJ) corpus and Mandarin Chinese AIShell-1 corpus \cite{bu2017}, following the data-preparation recipes in ESPNET toolkit \cite{watanabe2019}. Speech perturbation is only applied on the training set of AIShell-1 with factors of 0.9 and 1.1. The acoustic features are 80-dimensional filter-banks together with 3-dimensional pitch features, extracted from a 25 ms window with a 10 ms shift. Cepstral Mean and Variance Normalisation (CMVN) is performed on the training set, and the estimated parameters are also applied to the development and test sets. In terms of output labels, WSJ and AiShell-1 have 52 and 4231 classes, respectively.

The online Transformer model adopts a chunkwise self-attention encoder (chunk-SAE) as presented in \cite{miao2020b}. Non-overlapping chunks with length $N_c$ are spliced from the original utterance so that they could be sequentially fed into the model. To involve context information, $N_l$ left frames and $N_r$ right frames are complemented to each chunk, while only the central $N_c$ outputs are used by the SAD layers. Accordingly, the latency caused in chunk-SAE has been restricted to $N_r$.

A similar Transformer architecture is adopted for both the tasks, which is composed of a 12-layer chunk-SAE and a 6-layer SAD. Two identical CNN layers are used before the chunk-SAE, with each having 256 kernels of size $3\times3$ with a stride of $2\times2$ that reduces the frame rate by 2-fold. The chunk sizes of the chunk-SAE are $N_c=N_l=N_r=64$. The dimensions of all self-attention and DACS sub-layers plus the SAD's embedding layer are set to 256, and 4 heads are used in the multi-head mechanism. The position-wise feed-forward layers have 2048 units.

For training, we adopt CTC loss for performing multi-objective learning with the weight $\lambda_{CTC}=0.3$. Label smoothing is applied to the cross-entropy loss of Transformer with the penalty factor of 0.1. The models are trained up to 100 epochs for WSJ and 50 epochs for AIShell-1 using Adam optimizer \cite{kingma2014}, where the early-stopping criterion is imposed with the patience of 3 epochs. The learning rate follows Noam weight decay strategy \cite{vaswani2017}, which is initialized to 10.0 for WSJ and 1.0 for AIShell-1, while the warmup step is 25000 for both. To prevent over-fitting, dropout is applied to the self-attention and DACS sub-layers with the rate of 0.1. The mini-batch size is set to 32. Finally, the maximum look-ahead step $M$ is not applied during training, which is aimed at speeding up the matrix computation as discussed in Section 3. 

During inference, CTC scores are utilised to carry out joint online decoding with $\lambda_{CTC}=0.3$ for WSJ and $\lambda_{CTC}=0.5$ for AIShell-1. An external LM trained with the texts of training set is involved to rescore the beam search (beam width=10) hypotheses produced by the Transformer, where the LM for WSJ is an 1000-unit single layer Long Short Term Memory (LSTM) network, and for AIShell-1 it is a 650-unit 2-layer LSTM. Models of last 10 epochs are averaged and used for inference.

\makeatletter
\newcommand{\thickhline}{%
    \noalign {\ifnum 0=`}\fi \hrule height 1pt
    \futurelet \reserved@a \@xhline
}
\newcolumntype{"}{@{\hskip\tabcolsep\vrule width 1pt\hskip\tabcolsep}}
\makeatother

\begin{table}[t]
\centering
\caption{Word error rates (WERs) on WSJ.}
\begin{tabular}{llll}
\hline \thickhline
\multicolumn{1}{l}{Model}                          & \multicolumn{1}{c}{dev93} & \multicolumn{1}{c}{eval92} \\ \thickhline \hline
\multicolumn{3}{l}{Offline}                                                               \\ \hline
\multicolumn{1}{l}{BiLSTM \cite{karita2019} }                 & \multicolumn{1}{c}{-}      & \multicolumn{1}{c}{6.7}       \\ 
\multicolumn{1}{l}{Transformer \cite{karita2019}}            & \multicolumn{1}{c}{-}      & \multicolumn{1}{c}{4.9}       \\ \thickhline \hline
\multicolumn{3}{l}{Online}  \\ \hline
\multicolumn{1}{l}{{\begin{tabular}[c]{@{}l@{}}MoChA Transformer\\ \hspace{0.2cm} +SpecAug \cite{tsunoo2019}\end{tabular}}}  & \multicolumn{1}{c}{-}      & \multicolumn{1}{c}{6.6}  \\
\multicolumn{1}{l}{DACS Transformer (proposed)}       & \multicolumn{1}{c}{\textbf{8.9}}      & \multicolumn{1}{c}{\textbf{5.5}}       \\ \hline \thickhline
\end{tabular}
\end{table}

\begin{table}[t]
\centering
\caption{Character error rates (CERs) on AIShell-1.}
\begin{tabular}{llll}
\hline \thickhline
\multicolumn{1}{l}{Model}  &  & \multicolumn{1}{c}{dev} & \multicolumn{1}{c}{test} \\ \thickhline \hline
\multicolumn{3}{l}{Offline}                                                               \\ \hline
\multicolumn{1}{l}{BiLSTM \cite{karita2019} }    & & \multicolumn{1}{c}{-}      & \multicolumn{1}{c}{9.2}       \\ 
\multicolumn{1}{l}{SA-T \cite{tian2019}}     &  & \multicolumn{1}{c}{8.3}      & \multicolumn{1}{c}{9.3}       \\
\multicolumn{1}{l}{Transformer \cite{karita2019}}     & & \multicolumn{1}{c}{-}      & \multicolumn{1}{c}{6.7}       \\ \thickhline \hline
\multicolumn{3}{l}{Online}                                                                \\ \hline
\multicolumn{1}{l}{Chunk-flow SA-T \cite{tian2019}} & & \multicolumn{1}{c}{8.6}  & \multicolumn{1}{c}{9.8}    \\
\multicolumn{1}{l}{Sync-Transformer \cite{tian2020}} & & \multicolumn{1}{c}{7.9} & \multicolumn{1}{c}{8.9}     \\
\multicolumn{1}{l}{MTH-MoChA LSTM \cite{liu2020}}   & & \multicolumn{1}{c}{7.2} & \multicolumn{1}{c}{8.7}     \\
\multicolumn{1}{l}{\begin{tabular}[c]{@{}l@{}}MMA-MoChA Transformer\\ \hspace{0.2cm} (narrow-chunk) \cite{inaguma2020}\end{tabular}}       &  & \multicolumn{1}{c}{-}      & \multicolumn{1}{c}{7.5}       \\
\multicolumn{1}{l}{MoChA Transformer \cite{tsunoo2019}} & & \multicolumn{1}{c}{-}      & \multicolumn{1}{c}{9.7}       \\
\multicolumn{1}{l}{sMoChA Transformer }      &  & \multicolumn{1}{c}{7.9}      & \multicolumn{1}{c}{9.5}       \\
\multicolumn{1}{l}{MTA Transformer}               &  & \multicolumn{1}{c}{9.0}      & \multicolumn{1}{c}{9.7}       \\
\multicolumn{1}{l}{DACS Transformer (proposed)}    & & \multicolumn{1}{c}{\textbf{6.5}}      & \multicolumn{1}{c}{\textbf{7.1}}       \\ \hline \thickhline
\end{tabular}
\end{table}

\begin{table}[t]
\centering
\caption{WERs on WSJ and CERs on AIShell-1 with different maximum look-ahead steps $M$.}
\begin{tabular}{lllll} 
\hline \thickhline
\multirow{2}{*}{M}      & \multicolumn{2}{c}{WSJ}   & \multicolumn{2}{c}{AIShell-1}     \\ \cline{2-5} 
                        & \multicolumn{1}{c}{dev93}    & \multicolumn{1}{c}{eval92} & \multicolumn{1}{c}{dev}    & \multicolumn{1}{c}{test} \\ \thickhline
\multicolumn{1}{c}{$\infty$} & \multicolumn{1}{c}{8.4}  & \multicolumn{1}{c}{5.5} & \multicolumn{1}{c}{6.5} & \multicolumn{1}{c}{7.4}  \\ \hline
\multicolumn{1}{c}{16} & \multicolumn{1}{c}{8.9}  & \multicolumn{1}{c}{5.6} & \multicolumn{1}{c}{6.5}  & \multicolumn{1}{c}{7.4} \\
\multicolumn{1}{c}{14} & \multicolumn{1}{c}{8.9}  & \multicolumn{1}{c}{5.6} & \multicolumn{1}{c}{6.5}  & \multicolumn{1}{c}{7.4} \\
\multicolumn{1}{c}{12} & \multicolumn{1}{c}{8.8}  & \multicolumn{1}{c}{5.5} & \multicolumn{1}{c}{6.6}  & \multicolumn{1}{c}{7.5} \\
\multicolumn{1}{c}{10} & \multicolumn{1}{c}{8.9}  & \multicolumn{1}{c}{5.5} & \multicolumn{1}{c}{6.6}  & \multicolumn{1}{c}{8.1} \\
\multicolumn{1}{c}{8}  & \multicolumn{1}{c}{9.1}  & \multicolumn{1}{c}{5.6} & \multicolumn{1}{c}{7.4}  & \multicolumn{1}{c}{11.2} \\
\multicolumn{1}{c}{6}  & \multicolumn{1}{c}{9.6}  & \multicolumn{1}{c}{5.9} & \multicolumn{1}{c}{15.2}  & \multicolumn{1}{c}{24.1} \\
\multicolumn{1}{c}{4}  & \multicolumn{1}{c}{11.2}  & \multicolumn{1}{c}{6.9} & \multicolumn{1}{c}{51.6}  & \multicolumn{1}{c}{56.8} \\
\thickhline
\end{tabular}
\end{table}

\begin{table}[t]
\centering
\caption{Comparison of training and inference speed on AIShell-1 by different online Transformer ASR systems. The training times are given in minute.}
\begin{tabular}{llll} 
\hline \thickhline
\multicolumn{1}{l}{System}      & \multicolumn{1}{c}{train}   & \multicolumn{1}{c}{dev} & \multicolumn{1}{c}{test} \\ \thickhline
\multicolumn{1}{l}{offline}  & \multicolumn{1}{c}{36.7}  & \multicolumn{1}{c}{1.0}  & \multicolumn{1}{c}{1.0} \\ \hline
\multicolumn{1}{l}{MoChA}    & \multicolumn{1}{c}{106.7}  & \multicolumn{1}{c}{0.73}  & \multicolumn{1}{c}{0.75} \\
\multicolumn{1}{l}{sMoChA}   & \multicolumn{1}{c}{97.3}  & \multicolumn{1}{c}{0.72}  & \multicolumn{1}{c}{0.73} \\
\multicolumn{1}{l}{MTA}      & \multicolumn{1}{c}{93.5}  & \multicolumn{1}{c}{0.65}  & \multicolumn{1}{c}{0.65} \\
\multicolumn{1}{l}{DACS ($M=14$)}     & \multicolumn{1}{c}{\textbf{86.2}}  & \multicolumn{1}{c}{\textbf{0.62}}  & \multicolumn{1}{c}{\textbf{0.61}} \\
\thickhline
\end{tabular}
\end{table}

\subsection{Results}
\label{ssec:re}

Experimental results on WSJ and AIShell-1 are presented in Table 1 and 2, respectively. To compare the training and inference behaviours of different online Transformer ASR, MoChA, sMoChA and MTA based models are also generated on AIShell-1, using the same configurations described above. For fair comparisons, we only include the baseline systems that conduct CTC/attention joint training and incorporate a separate LM during inference. From the tables, one can observe that the proposed system achieves the best performance on both WSJ ($M$=10) and AIShell-1 ($M$=14) datasets when compared with the baseline systems. A relative gain of 16.7\% WER is obtained on the eval92 set of WSJ by DACS Transformer over the MoChA Transformer reported in \cite{tsunoo2019}. Also note that we do not perform SpecAug \cite{park2019} in our implementation for training on WSJ. On the other hand, DACS-based model achieves modest gains on AIShell-1 when compared with other online models. 

Table 3 summarises the effect of $M$ on the performance of Transformer using the DACS algorithm. In order to comply with the decoding behaviour of Bernoulli-based methods, DACS Transformer is first evaluated with $M=\infty$, which means that the accumulation of halting confidence is allowed to span all the encoder states for each output. The quasi-online system achieves 5.5\% WER on WSJ, and 7.4\% CER on AISHell-1 test sets respectively. Then we make the system work fully real-time by gradually decaying $M$ from 16, which corresponds to the latency caused by the chunk-SAE after subsampling. The smaller the value of $M$, the lesser the latency in recognition. One can observe that as the value of $M$ decreases beyond a certain value, the performance starts to degrade. It is also seen that the change in performance is more prominent on AIShell-1 than on WSJ when the value of $M$ is reduced from 10 to 8. As Chinese characters usually last for more frames than English letters, a smaller $M$ can impact the recognition performance. More strikingly, the online system achieves very close performance with the quasi-online counterpart on a wide range of latency settings.

\begin{figure}[t]

\begin{minipage}[t]{1.0\linewidth}
  \centering
  \centerline{\includegraphics[width=8.5cm]{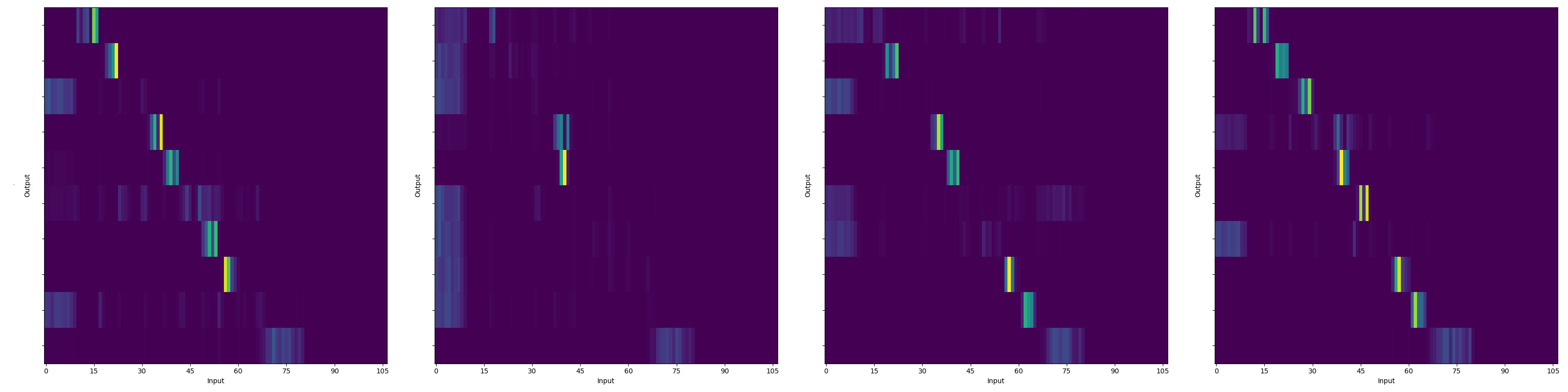}}
  \centerline{(a) Offline}\medskip
\end{minipage}
\begin{minipage}[t]{1.0\linewidth}
  \centering
  \centerline{\includegraphics[width=8.5cm]{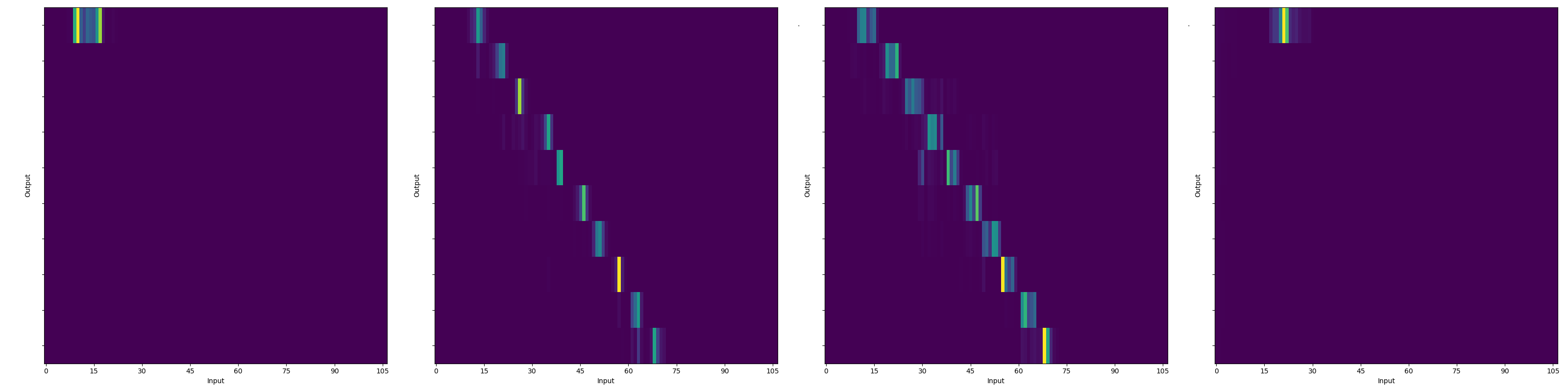}}
  \centerline{(b) MoChA}\medskip
\end{minipage}
\begin{minipage}[t]{1.0\linewidth}
  \centering
  \centerline{\includegraphics[width=8.5cm]{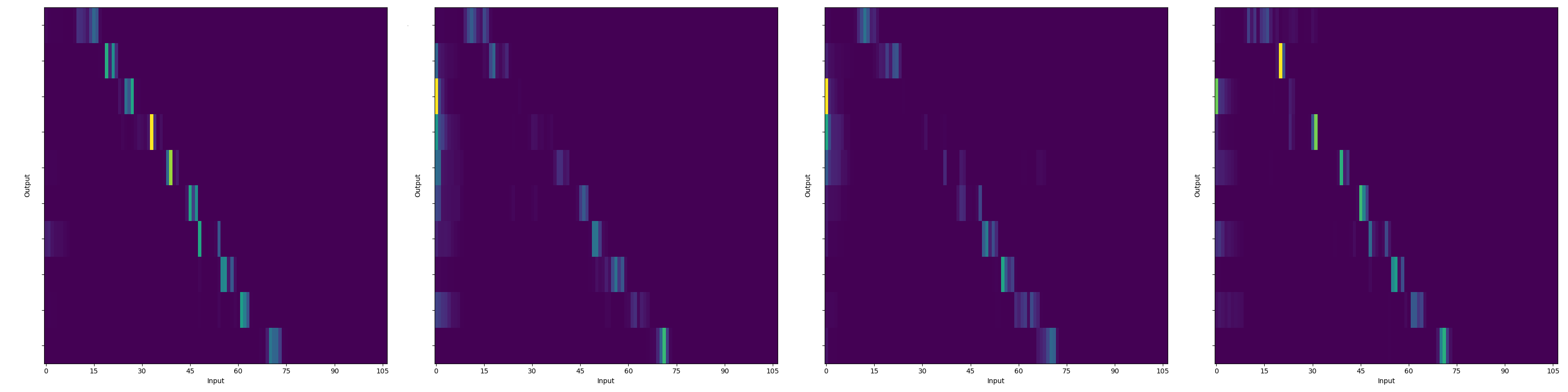}}
  \centerline{(b) sMoChA}\medskip
\end{minipage}
\begin{minipage}[t]{1.0\linewidth}
  \centering
  \centerline{\includegraphics[width=8.5cm]{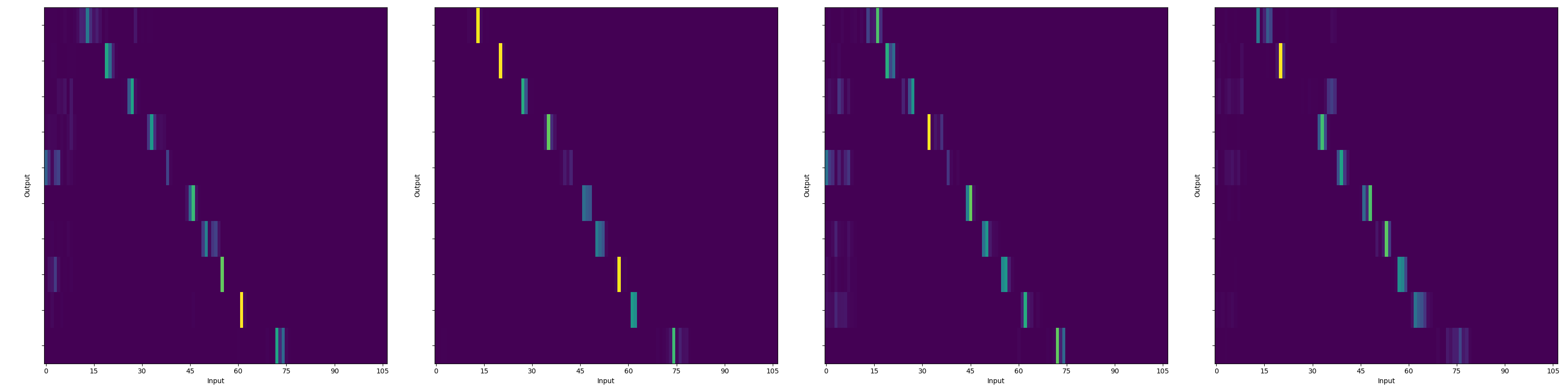}}
  \centerline{(d) MTA}\medskip
\end{minipage}
\begin{minipage}[t]{1.0\linewidth}
  \centering
  \centerline{\includegraphics[width=8.5cm]{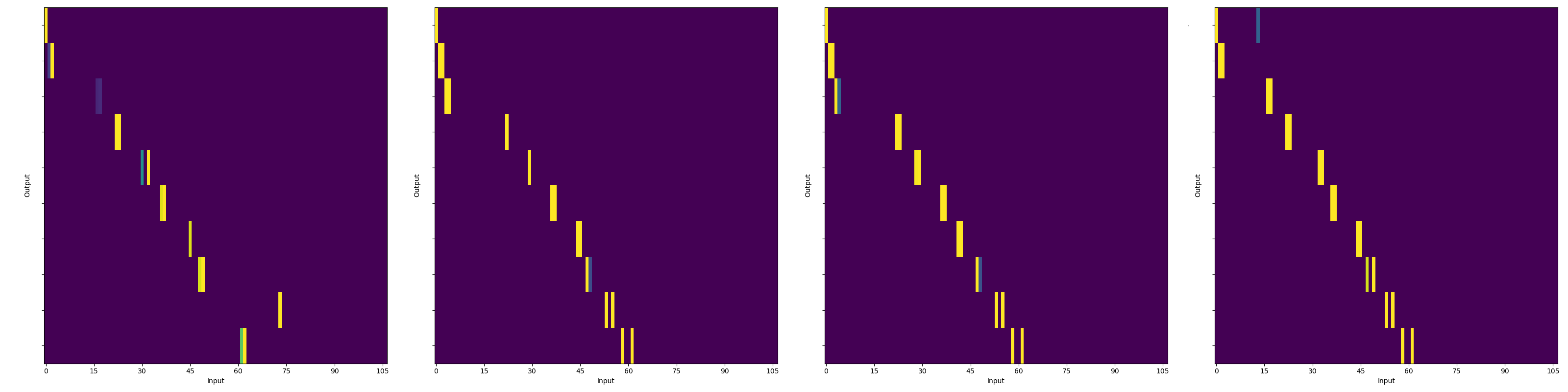}}
  \centerline{(e) DACS ($M=16$)}\medskip
\end{minipage}
\caption{Monotonic attention weights produced in the decoding process by offline and online Transformer ASR systems.}
\label{fig:res}
\end{figure}

\subsection{Comparisons of online Transformer ASR systems}
\label{ssec:behav}

Table 4 compares the speed of training and inference between DACS and Bernoulli-based Transformer ASR systems. The average training time per epoch on AIShell-1 is reported, where a single Geforce RTX 2080 Ti GPU card is used to train the models. From the table, one can notice that the training time decreases from MoChA to DACS, indicating the reduced complexity in calculation of the attention weights. As for the inference is concerned, the speed is measured in terms of the computational cost in the cross-attention sub-layers by defining a ratio:
\begin{equation}
r = \frac{\sum^{N_d}_{l=1} \sum^{H}_{h=1} \sum^{L}_{i=1} {s^{h,l}_i} }{(N_d \times H \times L) \times T},
\end{equation}
where $s^{h,l}_i$ is the number of computation steps performed at output step $i$ by $head_h$ in the $l^{th}$ SAD layer, and T denotes the number of encoding timesteps, with each aggregated for all $L$ output steps, $H$ heads and $N_d$ layers. The smaller the ratio is, the lesser cost is committed during inference. Note that in MoChA and sMoChA models, $s^{h,l}_i$ includes computation steps of both the first-pass monotonic attention and the second-pass soft attention. The value of $r$ is averaged over the whole development or test sets, and it can be observed that DACS yields the smallest computational cost than Bernoulli-based mechanisms.

Fig. 1 illustrates the MA weights of the top SAD layer to understand the speech-to-text-alignment of various online Transformer based approaches for ASR. The example utterance is taken from the AIShell-1 corpus. It is noticed that the attention weights of hard attention mechanisms like MTA are more concentrated on a narrow range of encoder states. On the contrary, MoChA and sMoChA tend to dilute the attentions as a result of applying eq. (10). From the perspective of magnitude, we find that there are plenty of output steps where attentions are missing or very subtle. Specifically, MoChA has two MA heads producing zeros alignments for most of the output steps. However, as reflected in Fig. 1(e), much sharper attention weights are obtained by DACS-based MA heads. This is possibly because the halting probabilities are calculated subject to neither the cumulative product like in eq. (7) and (12), nor the normalisation operations such as softmax. Another interesting behaviour is that the halting positions given by DACS are commonly earlier than the centres of attentions produced by other mechanisms. We believe that the truncation towards the halting confidence has pushed the encoding information backward in time, so as to minimise the loss of attention, a preferable property for online ASR.

\section{Conclusion}
\label{ssec:conclusion}

The paper presented a novel online attention mechanism, called DACS, that facilitated the use of Transformer for online ASR. At each decoder step, instead of operating on the entire input utterance, the DACS layer computed halting probabilities monotonically along the encoder states, and the computations were terminated once the accumulation of halting probabilities exceeded a certain threshold. The resulted probabilities were used in speech-to-text-alignment to predict the current output. Moreover, we also introduced a maximum look-ahead step to restrict the number of timesteps that the DACS layer could look at each output step, so as to reduce the latency in decoding. Experiments were presented on WSJ and AIShell-1 corpora to show the effectiveness of the proposed approach. The DACS based Transformer achieved 5.5\% WER on WSJ and 7.4\% CER on AIShell-1 test sets respectively, outperforming other online Transformer ASR systems and also further improving the speed for training and inference.



\bibliographystyle{IEEEbib}
\bibliography{refs}

\end{document}